\documentclass{kapproc} 

\usepackage{procps} 

\usepackage{graphicx}

\upperandlowercase

\normallatexbib 

\begin{document}

\input epsf.sty

\articletitle{Superconductivity in the background of 
two-dimensional stripe superstructure}


\author{Boris V. Fine}
\affil{Max Planck Institute for the Physics of Complex Systems\\
Noethnitzer Str. 38, 01187 Dresden, Germany}
\email{fine@mpipks-dresden.mpg.de}

\begin{abstract}
I propose a superconductivity model, which is based on the
assumption that stripes in high-$T_c$ cuprates (a) exist and (b) organize 
themselves in a two-dimen- sional superstructure. The model describes 
hole states, which are localized either inside the stripes or 
in the antiferromagnetic domains between the stripes. The  superconductivity
in this model emerges due to the interaction, which is, presumably,
mediated by the transverse fluctuations of stripes. 
The tunnelling density of states  obtained
from the mean field solution of the model 
is asymmetric with respect to the chemical potential, 
has Van Hove singularity identified as a superconducting peak,
and, in one of the model regimes, 
has linear functional form in the vicinity 
of the chemical potential.
The relation between the critical temperature and 
the zero-temperature superfluid density has ``fish-like''
form, which quantitatively resembles experimental data.
The superconducting order parameter obtained from this model 
has two components exhibiting non-trivial phase and sign change under
translations in real space.
\end{abstract}


If energetically {\it deep} stripes  absorbing most of the charge carriers
are present in a 
superconducting (SC) material, then  it is likely that the SC mechanism 
operating in this material would not be operational
without stripes. If one further accepts that deep 
stripes exist in the La$_{2-x}$Sr$_x$CuO$_4$ (LSCO) family
of high-$T_c$ cuprates, then it  implies that the SC mechanism in LSCO
is not operational without stripes. Finally, if one also assumes that the
SC mechanism is the same in all families of high-$T_c$
cuprates, then the unavoidable conclusion is that stripes, or, at
least strong local inhomogeneities exist in all families of high-$T_c$ cuprates
and play a crucial role in the mechanism of superconductivity.

\begin{figure} \setlength{\unitlength}{0.1cm}

\begin{picture}(100, 65) 
{ 
\put(-2, 0){ \epsfxsize= 2.3in \epsfbox{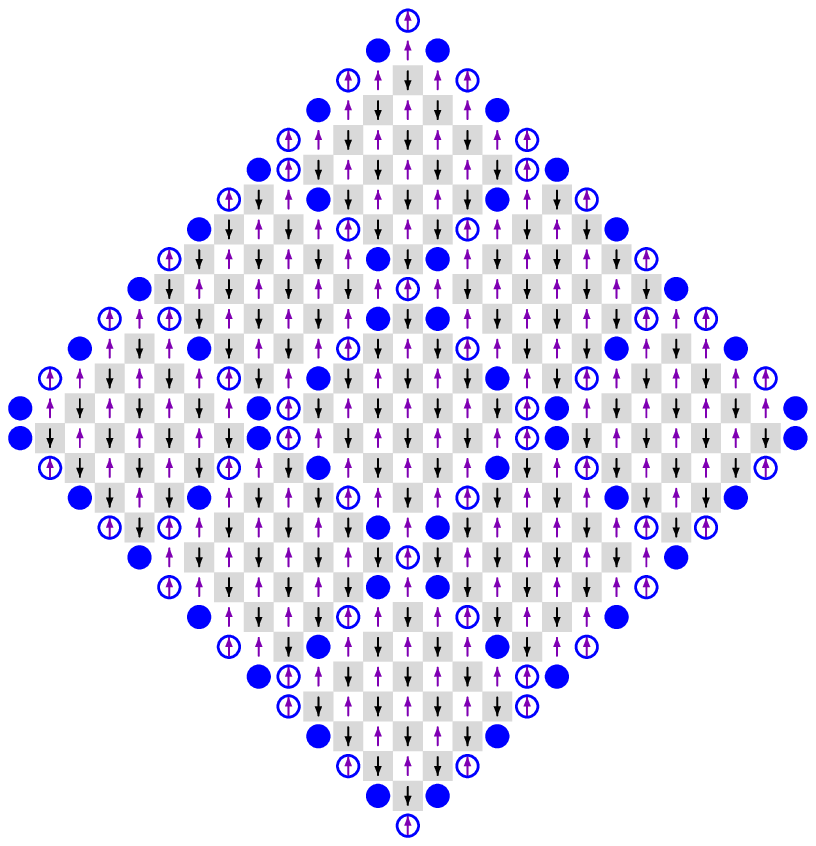} }
\put(62, 1){ \epsfxsize= 2.3in \epsfbox{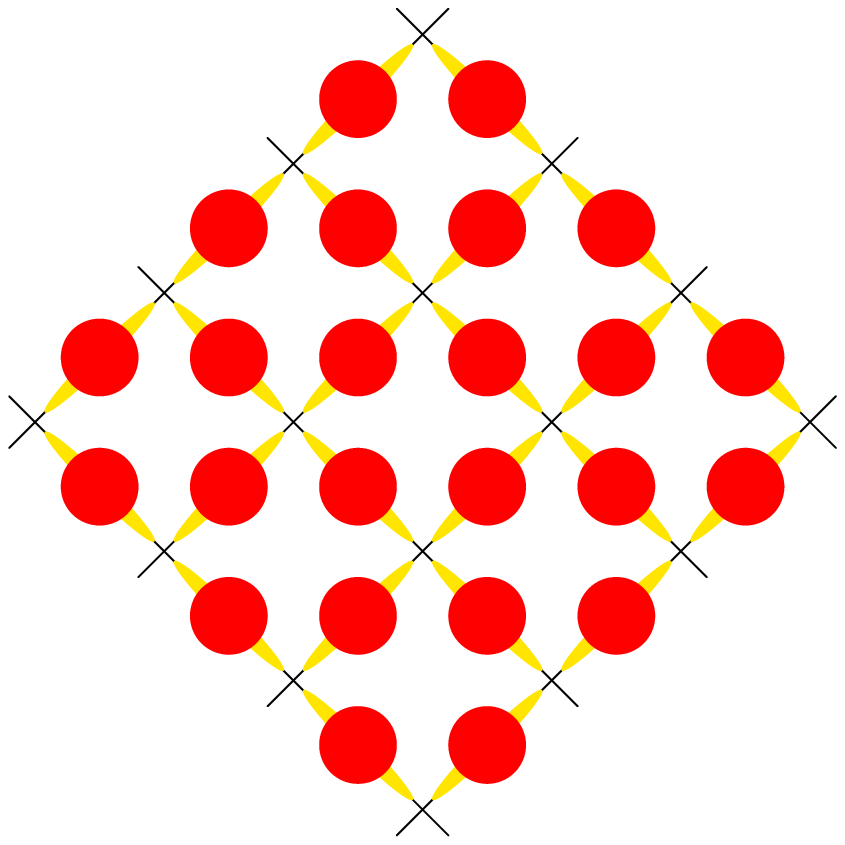} }
\put(3,50){(a)}
\put(67,50){(b)}
}
\end{picture} 
\caption{(a) 2D configuration of diagonal stripes.  
(b) Checkerboard pattern formed by b-states. 
Circles represent the regions, where
b-states are localized. 
} 
\label{fig-stripes} 
\end{figure}

In LSCO, the basic evidence of (dynamic) stripes comes in the form of the 
well-known four-fold splitting of magnetic $(\pi,\pi)$ peak, which is observed by
inelastic neutron scattering\cite{Yamada-etal}, 
and corroborated by the observation of
the elastic response with similar peak pattern 
in Nd-doped LSCO\cite{Tranquada-etal}.
The above four-fold splitting has been generally interpreted in the 
``stripe community'' as the evidence for two stripe domains, 
each characterized
by a one-dimensional  array of  stripes running along one of the principal 
lattice directions. This picture, however, runs into many difficulties, given
numerous manifestly two-dimensional (2D) properties of high-$T_c$ cuprates.
An alternative interpretation of the four-fold peak pattern, which has been
discussed in the literature (see, e.g., Ref.~\cite{SG}) 
but never pursued very far, 
would be based on 
the 2D arrangement of stripes shown in Fig.~\ref{fig-stripes}.
The purpose of the present work is to show that the above 2D picture 
is compatible with superconductivity in general, and 
with the phenomenology of high-$T_c$ cuprates in particular. 
This manuscript constitutes a compressed version of a longer paper 
(Ref.~\cite{Fine-hitc-condmat03}). 

The 2D stripe background shown in Fig.~\ref{fig-stripes} implies the 
existence of two kinds of hole states: a-states --- localized 
inside the antiferromagnetic (AF) domains, and b-states --- 
localized inside the stripes. The on-site energies associated with 
a- and b-states will be denoted as $\varepsilon_a$ and $\varepsilon_b$,
respectively. In underdoped cuprates, the expectation is that 
$\varepsilon_a > \varepsilon_b$. The typical value of the difference 
$\varepsilon_a - \varepsilon_b$ should then be identified with the pseudogap.
The stripe superstructure  should strongly suppress 
the transport of holes. Therefore, in the zeroth-order approximation,
it is reasonable to neglect the direct hopping between both a-states 
and b-states belonging to different units of the stripe
superstructure and also  exclude all interaction terms, 
which shift the center of mass of the hole subsystem.

Here, I introduce, perhaps, the most simple model, which satisfies the above
``selection rule''.
The model configuration  includes one a-state per AF domain, and two b-states 
per stripe element having opposite
orientations of spins.
(Stripe element is a piece of a stripe confined between two subsequent 
intersections with perpendicular stripes.) 
The model Hamiltonian is:
\begin{equation}
{\cal H} =\varepsilon_a \sum_i a_i^+ a_i \ + \ 
\varepsilon_b \sum_{i, j(i), \sigma}^{\eta_i= 1} b_{ij,\sigma}^+ b_{ij,\sigma} 
+ \ g \sum_{i, j(i)}^{\eta_i= 1} ( b_{ij,+}^+ b_{ij,-}^+ a_i a_j + \hbox{h. c.}),
\label{H}
\end{equation}
where single index $i$ or $j$ labels AF domains; 
notation $j(i)$ implies, that the $j$th AF domain is the nearest neighbor
of the $i$th domain; $a_i$ is the annihilation 
operator of a hole inside the $i$th AF domain;
$b_{ij,\sigma}$ is the annihilation operator of a hole inside 
the stripe element separating
the $i$th and the $j$th AF domains; $\sigma$ is the spin index, which
can have two values ``$+$'' or ``$-$''; 
$\varepsilon_a$ and $\varepsilon_b$
are position-independent 
on-site energies for a- and b-states, respectively, counted from the
chemical potential; and, finally, $g$ is the coupling constant.
The spin wave function of a-states  alternates 
together with the AF order parameter, i.e. a-states belonging to neighboring
AF domains always have opposite spins. The spins of a-states 
are tracked by index $\eta_i$, which 
can have values $1$ or $-1$. The supercells corresponding
to $\eta_i =1$ and $\eta_i =-1$ are to be called ``even'' and ``odd,'' 
respectively. 
The sum superscript ``$\eta_i =1$'' in Eq.(\ref{H}) 
indicates that the summation extends only
over even supercells.
Each transition corresponding to the interaction term in the
Hamiltonian~(\ref{H}) can be described either as  
``two
holes from the opposite sides of a given stripe element 
hopping simultaneously into that element'', or as the
reverse process.  
This kind of interaction is, presumably, mediated by the
transverse fluctuations of stripe elements.

The mean-field solution of the above model consists of 
(i) making the Fourier transform of even $a$-operators
and odd $a$-operators {\it separately}, which gives,
respectively, $a_{\hbox{e}}({\mathbf{k}})$  and 
$a_{\hbox{o}}({\mathbf{k}})$;
and (ii) introducing the following Bogoliubov transformations:
\begin{eqnarray}
a_{\hbox{e}}({\mathbf{k}}) &=& u({\mathbf{k}}) A_{\hbox{e}}({\mathbf{k}}) 
+ v({\mathbf{k}}) e^{i \phi_a({\mathbf{k}})} 
A^+_{\hbox{o}}(-{\mathbf{k}}) , 
\label{aeA}
\\
a_{\hbox{o}}(-{\mathbf{k}}) &=& u({\mathbf{k}}) 
A_{\hbox{o}}(-{\mathbf{k}}) 
- v({\mathbf{k}}) e^{i \phi_a({\mathbf{k}})} 
A^+_{\hbox{e}}({\mathbf{k}}),  
\label{aoA}
\end{eqnarray}
\begin{eqnarray}
b_{ij+} = s B_{ij+} \  + \ w e^{i \varphi_{ij}} B^+_{ij-};
\label{bB+2}
\\
b_{ij-} = s B_{ij-} \ - \ w e^{i \varphi_{ij}} B^+_{ij+},
\label{bB-2}
\end{eqnarray}
where  
$A_{\hbox{e}}({\mathbf{k}})$,  $A_{\hbox{o}}({\mathbf{k}})$ and
$B_{ij,\sigma}$ 
are the annihilation operators of new Bogoliubov quasiparticles;
$\phi_a({\mathbf{k}})$ and $\varphi_{ij}$ are the phases of these 
transformations; and
$u({\mathbf{k}})$, $v({\mathbf{k}})$, $s$ and $w$ are the real
numbers 
obeying the following
normalization conditions:
$u^2({\mathbf{k}}) + v^2({\mathbf{k}}) = 1 $; 
$s^2 + w^2 = 1$.
Phases $\varphi_{ij} \equiv \varphi({\mathbf{r}}_j - {\mathbf{r}}_i)$ 
are  chosen to be the same
for all translationally equivalent stripe elements.
Four kinds of translationally non-equivalent stripe elements
correspond to four possible even-to-odd 
nearest neighbor translation vectors 
${\mathbf{R}}_1 =  (1, 1)\ {l / \sqrt{2}}$; 
${\mathbf{R}}_2 =  (-1, 1)\ {l / \sqrt{2}}$;
${\mathbf{R}}_3 =  (-1, -1)\ {l / \sqrt{2}}$; and 
${\mathbf{R}}_4 =  (1, -1)\ {l / \sqrt{2}}$. (Here $l$ is 
the length of a stripe element.) Correspondingly,
there exist four independent phases 
$\varphi_{\alpha} = \varphi({\mathbf{R}}_{\alpha})$.

Below I consider two most promising cases: Case I --- characterized by 
\mbox{$\varepsilon_b = 0$}, and Case II --- characterized by 
$\varepsilon_a = 0$. In Case I, 
the standard variational scheme leads to the following equations 
for the critical temperature~($T_c$):
\begin{equation}
T_c = { g^2 \left[\hbox{exp}\left({|\varepsilon_a| \over T_c}\right) - 1\right] \over 
8 |\varepsilon_a| \left[\hbox{exp}\left({|\varepsilon_a| \over T_c}\right) + 1\right]},
\label{Tceq}
\end{equation}
and for the zero-temperature energies of A- and B- quasiparticles :
\begin{equation}
\varepsilon_A({\mathbf{k}}) = \sqrt{\varepsilon_a^2 + 
{1 \over 4} g^2  |V({\mathbf{k}})|^2
},
\label{epsAkI}
\end{equation}
\begin{equation}
\varepsilon_B = {g^2  \over 8N}
\sum_{{\mathbf{k}}} { |V({\mathbf{k}})|^2
\over \varepsilon_A({\mathbf{k}})},
\label{epsBI}
\end{equation}
where N is the total number of supercells, and 
$ V({\mathbf{k}})  =   \sum_{\alpha} 
e^{- i \varphi_{\alpha} - i {\mathbf{k}} {\mathbf{R}}_{\alpha}} $.
The four phases $\varphi_{\alpha}$  are only constrained
by condition $
{(\varphi_2 + \varphi_4 -\varphi_1 - \varphi_3)/ 2} =
{\pi / 2} + \pi n,
$,
where $n$ is an integer number.
The density of  B-states thus
consists of two symmetric $\delta$-function
peaks located at $\pm \varepsilon_B$. The density of A-states, 
is continuous but
asymmetric with respect to the chemical potential. 
It has Van Hove singularities  at 
$ \varepsilon_{A0} = \pm \sqrt{ \varepsilon_a^2 +   g^2 }$,
and a gap extending between $-\varepsilon_a$ and $\varepsilon_a$.

In Case II, the analogous results are:
\begin{equation}
T_c = { g^2 \left[\hbox{exp}\left({|\varepsilon_b| \over T_c}\right) - 1\right] \over 
8 |\varepsilon_b| \left[\hbox{exp}\left({|\varepsilon_b| \over T_c}\right) + 1\right]}.
\label{TceqII}
\end{equation}
\begin{eqnarray}
\varepsilon_A({\mathbf{k}}) &=&
 {g^2 \ |V({\mathbf{k}})| \ C_{a0}  \over 8 \varepsilon_B},
\label{epsAkII}
\\
\varepsilon_B &=& \sqrt{\varepsilon_b^2 + 
\ g^2 \ C_{a0}^2/16
},
\label{epsBII}
\end{eqnarray}
where
$V({\mathbf{k}})$ is the same as in Case I, and 
$
C_{a0} \equiv {1 \over N} \sum_{{\mathbf{k}}} |V({\mathbf{k}})|   
= 0.958 
$... .
In this case, the density of A-states is symmetric and has 
Van Hove singularities
at  $\varepsilon_{A0} = \pm {g^2 C_{a0}  \over 4 \varepsilon_B}$, 
while the $\delta$-peaks, 
corresponding to B-states are asymmetric.
Unlike the result for Case I, the density of A-states in Case II has no gap
around the chemical potential. Instead, it equals zero at the chemical 
potential and then increases linearly.

In order to interpret the experimental tunnelling data, it is necessary to 
assume, that the observed spectra are those of A-states, which means that
$\varepsilon_{A0}$ corresponds to the energy of the experimentally
observed SC peak.
The B-states are, perhaps, more difficult to observe. However, 
B-states (in the SC state), or b-states (in the normal state)
form a checkerboard pattern shown in  Fig.~\ref{fig-stripes}(b).
Therefore, they may be responsible for 
the checkerboard patterns seen by scanning 
tunnelling microscopy\cite{Hoffman-etal-02,Howald-etal,Vershinin-etal04}.

The superfluid properties of this model are unusual because of the unusual form
of the current operator. Fundamental to this model 
is the internal current operator,
which describes the particle flow between a- and b-states.
For the $i$th supercell, the internal current operator 
can be obtained as follows:
\begin{equation}
J_{ab(i)} \equiv - {d \over dt} (a_i^+ a_i) = 
- {i g \over \hbar} \sum_{j(i)} 
( b_{ij,+}^+ b_{ij,-}^+ a_i a_j -  \hbox{h.c.}) .
\label{Jabi}
\end{equation} 
Operator  (\ref{Jabi})
sums over four possible transitions, each transferring
a hole from $i$th AF domain to one of the four surrounding stripe elements.
When the direction
of each of the above transitions is taken into account,
the following expression for the translational current operator can be obtained:
\begin{equation}
{\mathbf{J}}^t_i
= - {i g \over 2 \hbar} \sum_{j(i)} \ 
\hat{{\mathbf{n}}}_{ij} \  \left( b_{ij,+}^+ b_{ij,-}^+ a_i a_j - \hbox{h.c.} \right) ,
\label{Jti}
\end{equation}
where $\hat{{\mathbf{n}}}_{ij}$ is the unit vector in the direction from the 
$i$th to the $j$th supercell.

The internal {\it supercurrent} corresponding to operator~(\ref{Jabi})
emerges, when the SC solution is modified by adding an extra 
phase $\phi_{ab}$ to $\varphi_{ij}$ in the Bogoliubov transformation for
b-states. 
If the phase $\phi_{ab}$ is the same for all stripe elements,
then the translational supercurrent equals zero. 
However, when  $\phi_{ab}$
has a weak position dependence, the zero-temperature 
density of translational supercurrent can be expressed as:
\begin{equation}
{\mathbf{j}} = {e \over l z_0}  \langle {\mathbf{J}}^t_i \rangle
= S_{\phi} \nabla \phi_{ab},
\label{j}
\end{equation}
where, in Case I,   
\begin{equation}
S_{\phi} = {e g^2  \over 16 N \hbar z_0} 
\sum_{\mathbf{k}} 
{ |V({\mathbf{k}})|^2 \over 
\varepsilon_A({\mathbf{k}}) };
\label{SphiI}
\end{equation}
and, in Case II, 
\begin{equation}
S_{\phi} = {e g^2  C_{a0}^2 \over 32 \ \hbar \ z_0 \ \varepsilon_B }.
\label{SphiII}
\end{equation}
Here, $z_0$ is the transverse distance per one SC plane, 
and $S_{\phi}$ is the SC phase stiffness 
(frequently referred to as superfluid density).

\begin{figure} \setlength{\unitlength}{0.1cm}

\begin{picture}(100, 36) 
{ 
\put(5, 0) { \epsfxsize= 1.4in \epsfbox{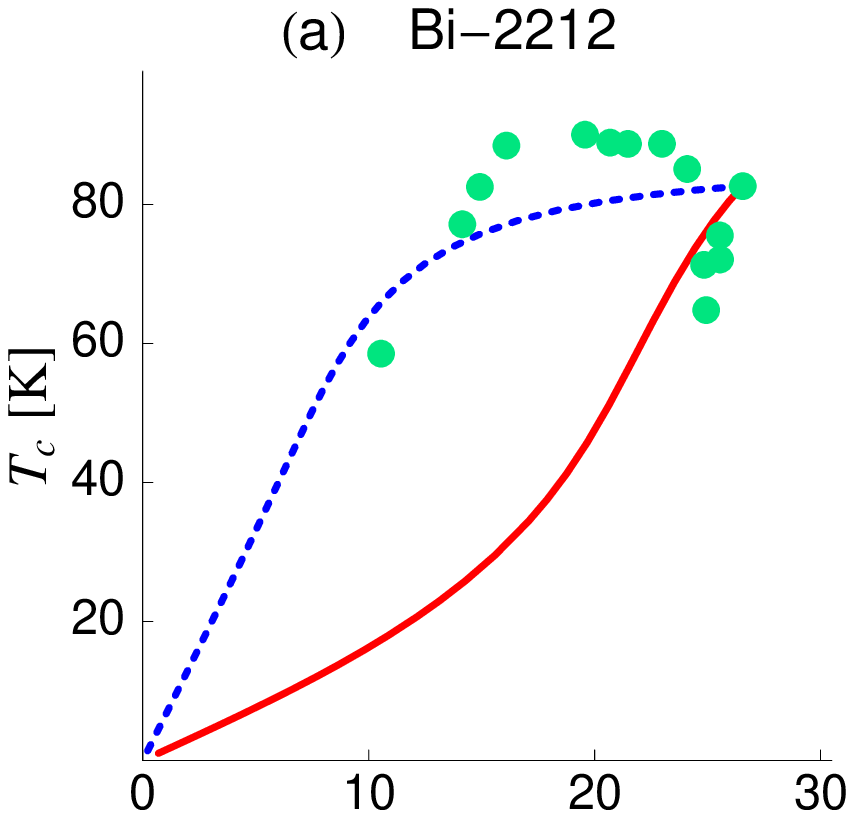}  }
\put(40, -1) { \epsfxsize= 1.4in \epsfbox{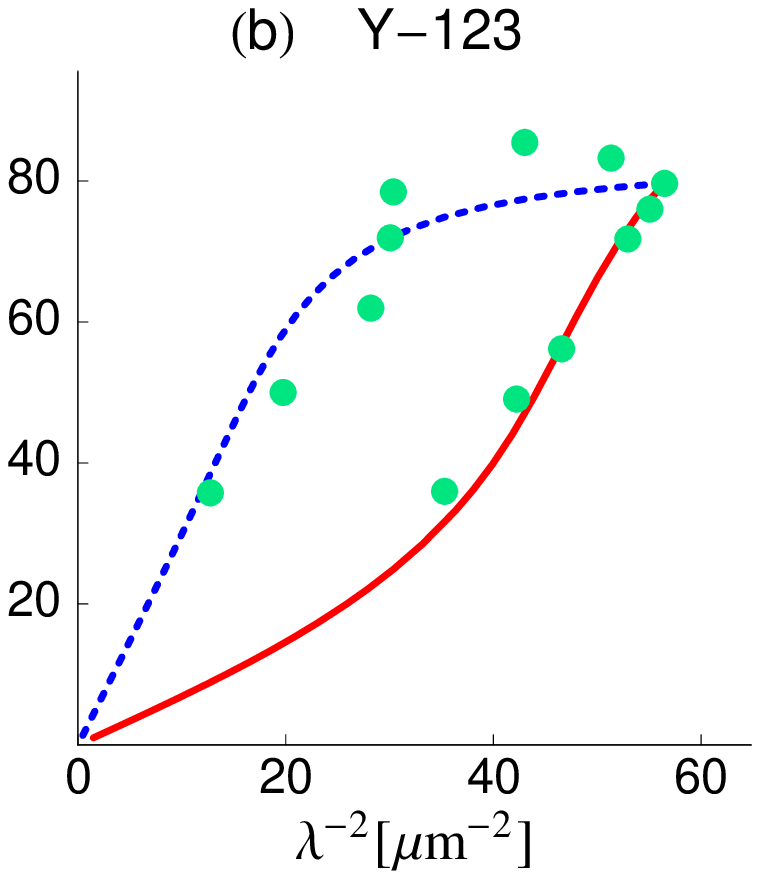}  }
\put(75, 0) { \epsfxsize= 1.4in \epsfbox{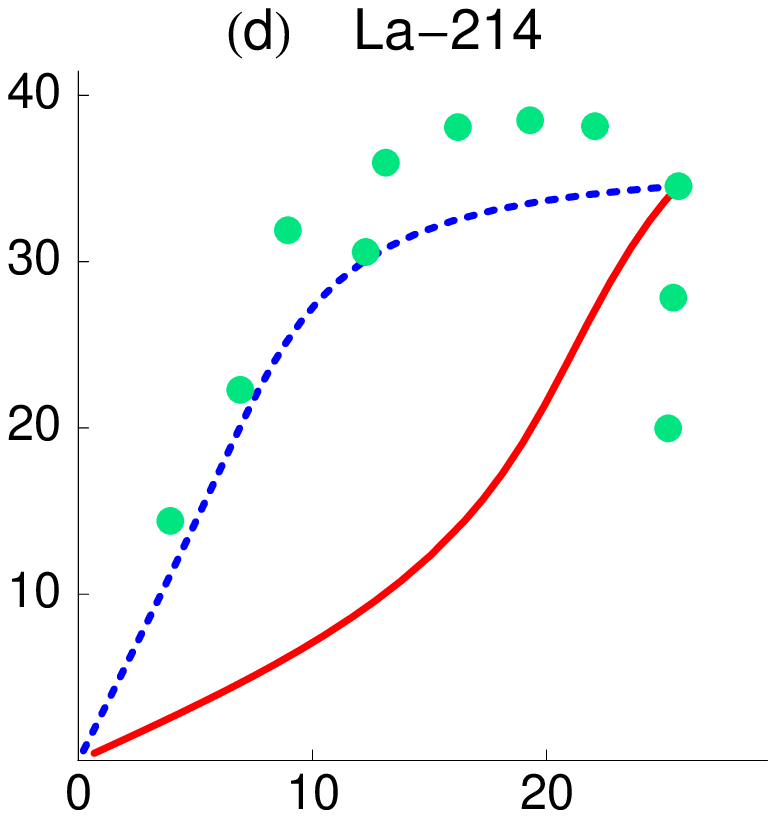} } 
}
\end{picture} 
\caption[]{ $T_c$ vs. $S_{\phi}$  
(presented as $1/\lambda^2$, where $\lambda$ is the in-plane penetration depth
of magnetic field): solid line --- Case I;  dashed line --- Case~II, 
circles ---
experimental points from Ref.~\protect\cite{Tallon-etal03}. Each plot
contains the same pair of  theoretical curves, which, initially,
were calculated in arbitrary units
as described in the text, and then  rescaled in such a way that 
the experimental critical points (evident in every plot) match 
the theoretical critical point corresponding to 
$\varepsilon_a = \varepsilon_b = 0$.
} 
\label{fig-sfexp} 
\end{figure}

As the doping concentration changes, the value of 
$\varepsilon_a - \varepsilon_b$ (characterizing the pseudogap)
should, in relative terms, change stronger than the coupling constant $g$. 
Therefore, an approximate relation between $T_c$ and $S_{\phi}$
within one family of high-$T_c$ cuprates can be obtained by
fixing the value of $g$ and then calculating  
$T_c$ and $S_{\phi}$ as  functions of
$\varepsilon_a$ (in Case I), or $\varepsilon_b$ (in Case II). 
The resulting theoretical 
relation is compared with experiments in Fig.~\ref{fig-sfexp}.

In conclusion, I have shown  that superconductivity is 
compatible with 2D stripe superstructure. The specific model presented in this
work has the following qualitative features resembling the phenomenology 
of high-$T_c$ cuprates:
(i) emergence of the quasiparticle coherence in $k$-space only 
at temperatures below $T_c$ (see $\varepsilon_A(k)$);
(ii) linear density of states in the vicinity of
the chemical potential (in Case II);
(iii) asymmetry in the tunnelling characteristics;
(iv) Van Hove singularity in the tunnelling density of states 
($\varepsilon_{A0}$);
(v) real space checkerboard pattern in the density of states;
(vi) low superfluid density having universal ``fish-like'' dependence
on $T_c$.
Although not discussed in this paper,   the Bogoliubov 
transformations (\ref{aeA}-\ref{bB-2}) imply 
a very unconventional symmetry of the 
SC order parameter, which, in particular, includes the sign change
of at least one of the two SC components under translations in real space
\cite{Fine-hitc-condmat03}. A similar prediction has also been made
by Ashkenazi in Ref.~\cite{Ashkenazi}.

{
\bibliographystyle{unsrt}
\chapbblname{miami}
\chapbibliography{hitc}

}

\end{document}